\documentclass[showpacs,preprintnumbers,amsmath,amssymb]{revtex4}
\usepackage{graphicx}% Include figure files
\usepackage{array}

\input epsf
\usepackage{dcolumn}% Align table columns on decimal point
\usepackage{bm}% bold math

\begin{document}

%\preprint{Theoretical and Mathematical Physics, 139 (3): 787-800 (2004)}

\title{Astronomical bounds on future big freeze singularity}% Force line breaks with \\
\author{Artyom V. Yurov}
\email{artyom_yurov@mail.ru}
%\altaffiliation[Also at ]
{%
\affiliation{%
I. Kant Russian State University, Theoretical Physics Department,
 Al.Nevsky St. 14, Kaliningrad 236041, Russia
\\
 }

\author{Artyom V. Astashenok}
\email{artyom.art@gmail.com}
%\altaffiliation[Also at ]
{%
\affiliation{%
I. Kant Russian State University, Theoretical Physics Department,
 Al.Nevsky St. 14, Kaliningrad 236041, Russia
\\
}
\author{Pedro F. Gonzalez-Diaz}
\email{p.gonzalezdiaz@imaff.cfmac.csic.es}
%\altaffiliation[Also at ]
{%
\affiliation{%
Colina de los Chopos, IMAFF, CSIC, Servano 121, Madrid 28006,
Spain
\\
}

\begin{abstract}
Recently it was found that dark energy in the form of phantom
generalized Chaplygin gas may lead to a new form of the cosmic
doomsday, the big freeze singularity.  Like the big rip singularity,
the big freeze singularity would also take place at a finite future
cosmic time, but unlike the big rip singularity it happens for a
finite scale factor.Our goal is to test if a universe filled with
phantom generalized Chaplygin gas can conform to the data of
astronomical observations. We shall see that if the universe is only
filled with generalized phantom Chaplygin gas with equation of state
$p=-c^2s^2/\rho^{\alpha}$ with $\alpha<-1$, then such a model cannot
be matched to the data of astronomical observations. To construct
matched models one actually need to add dark matter. This procedure
results in cosmological scenarios which do not contradict the data
of astronomical observations and allows one to estimate how long we
are now from the future big freeze doomsday.
\end{abstract}


\maketitle

\linespread{2}

\section{Introduction}

Till recently there were only two types of singularities in
cosmology: The big bang and the big crunch. The latter singularity
appeared in the universe if it would go through a contracting stage
which is possible both in closed universes (filled with baryon
matter and radiation) and in open universes filled with scalar field
with negative potential \cite{negav}.

This picture changed dramatically with the coming of  the dark
energy revolution. Now we know about many new scenarios for the end
of the universe among which there is the  big rip singularity
\cite{Caldwell} and the sudden future singularity \cite{sfs}.
Another way to obtain a cosmological doomsday was suggested in
\cite{page-2}. According to this scenario our vacuum should be
rather unstable and should decay within 20 Gyr (which is possible if
the gravitino is superheavy).

A new type of future singularity, the so-called big freeze
singularity, has been quite recently considered \cite{bf}. It is a
type III singularity in the notation of Ref. \cite{Noj}. Such a type
of doomsday is possible if the universe is filled with phantom
generalized Chaplygin gas  (which was originally introduced in
\cite{5}, \cite{10}, \cite{6}). Like the big rip singularity, this
singularity would also take place at a finite future cosmic time,
but unlike the big rip singularity, it happened for a finite scale
factor. This feature results in very interesting conclusions some of
which are discussed in this paper.

The aim of this article is to find what bounds on the future
lifetime of the universe filled with phantom generalized Chaplygin
gas  can be placed by current cosmological observations. As we shall
see in Sec.2 and Sec.3,  if the universe is only filled with
generalized phantom Chaplygin gas with equation of state
$p=-c^2s^2/\rho^{\alpha}$ with $\alpha<-1$, then such a model cannot
match the data of astronomical observations. To construct models
compatible with observations one need to add dark matter which
allows one to obtain a class of good models where the total lifetime
of the universe can be calculated. It is interesting to note that
whereas the present age of the universe $t_0$ is weakly depending on
the parameter of the equation of state $w$ and the Chaplygin
parameter $\alpha$, the total lifetime of the universe $t_f$ and the
time to be elapsed from now to the final big freeze singularity
($t_f-t_0$) are very sensitive to these parameters. For example, if
$\alpha=-2$ then for the flat universe with $w=-1.01$ we have
$t_f-t_0=10.3$ Gyr, $t_0=14.4$ Gyr, whereas for the case with
$w=-1.05$ we have $t_f-t_0=6.4$ Gyr and $t_0=13.8$ Gyr.

\section{The model}
Let us consider a universe filled with dark matter ($p=0$) and a
generalized Chaplygin gas with equation of state
$p=-c^{2}s^{2}\rho^{1+\epsilon/3}=-c^{2}\beta^{2}a_{f}^{\epsilon}\rho^{1+\epsilon/3}$.
The the Friedmann equations would read

\begin{equation}\label{fr1}
\frac{\dot{a}^{2}}{a^{2}}=\frac{8\pi
G}{3}\left(\rho+\frac{c_{d}^{2}}{a^{3}}\right)-\frac{kc^{2}}{a^{2}},
\end{equation}

\begin{equation}\label{fr2}
\frac{\ddot{a}}{a}=-\frac{4\pi
G}{3}\left(\rho+\frac{3p}{c^{2}}+\frac{c_{d}^{2}}{a^{3}}\right),
\end{equation}

where we have denoted
$\rho=\beta^{-6/\epsilon}(a_{f}^{\epsilon}-a^{\epsilon})^{-3/\epsilon}$
and $p=-c^{2}\beta^{2}a_{f}^{\epsilon}\rho^{1+\epsilon/3}$ for,
respectively, the energy density and pressure of the generalized
Chaplygin gas.

Observations demand that

\begin{equation}\label{accel}
\frac{\ddot{a}_{0}}{a_{0}}=k_{0}H_{0}^{2},
\end{equation}

where $k_{0}\approx 0.7$. Using (\ref{fr1},\ref{fr2},\ref{accel})
we get

\begin{equation}\label{pchapl}
p_{0}=-\frac{c^{2}}{8\pi
G}\left((2k_{0}+1)H_{0}^{2}+\frac{kc^{2}}{a_{0}^{2}}\right).
\end{equation}

However,
$p_{0}=-c^{2}a_{f}^{\epsilon}\beta^{-6/\epsilon}(a_{f}^{\epsilon}-a_{0}^{\epsilon})^{-1-3/\epsilon}$,
therefore we can find the unknown constant $\beta$ given by
\begin{equation}\label{beta}
\beta^{-6/\epsilon}=\frac{a_{0}^{3}\left(N_{0}-1\right)^{1+3/\epsilon}}{8\pi
GN_{0}}\left((2k_{0}+1)H_{0}^{2}+\frac{kc^{2}}{a_{0}^{2}}\right),
\end{equation}

where $N_{0}=\frac{a_{f}^{\epsilon}}{a_{0}^{\epsilon}}$. The current
parameter of the equation of state is
$w_{0}=\frac{p_{0}}{c^{2}\rho_{0}}$, so that
\begin{equation}\label{w0}
    w_{0}=\frac{p_{0}}{c^{2}\rho_{0}}=-\beta^{2}N_{0}a_{0}^{3}\rho_{0}^{\epsilon/3}
    =-\frac{N_{0}}{N_{0}-1}
\end{equation}

Finally, we have for the current vacuum energy density
\begin{equation}\label{rho0}
    \rho_{0}=\frac{N_{0}-1}{8\pi
    GN_{0}}\left(\frac{kc^{2}}{a_{0}^{2}}+\left(2k_{0}+1\right)H_{0}^{2}\right)
\end{equation}

Using (\ref{fr1}) and (\ref{rho0}) we can get $c_{d}^{2}$:
\begin{equation}\label{cd2}
    c_{d}^{2}=\frac{H^{2}_{0}a_{0}^{3}}{4\pi
    Gw_{0}}\left(\frac{3w_{0}+1}{2}\left(\frac{kc^{2}}{H_{0}^{2}a_{0}^{2}}+1\right)+k_{0}\right)
\end{equation}

and

\begin{equation}\label{omega0}
    \Omega_{0}\equiv\frac{\rho_{0}}{\rho_{c}}=\frac{N_{0}-1}{3N_{0}}\left(\frac{kc^{2}}{a_{0}^{2}H_{0}^{2}}+2k_{0}+1\right).
\end{equation}

Using (\ref{w0}) we can derive from (\ref{omega0}) the second
formula for $\Omega_{0}$
\begin{equation}\label{Omega01}
    \Omega_{0}=-\frac{1}{3w_{0}}\left(\frac{kc^{2}}{a_{0}^{2}H_{0}^{2}}+2k_{0}+1\right)
\end{equation}
We introduce now the parameter
$\Omega_{d,0}=\frac{\rho_{d,0}}{\rho_{c}}$. It is obvious that
$$\Omega_{d,0}=1+\frac{kc^{2}}{H_{0}a_{0}^{2}}-\Omega_{0}.$$
Because $w_{0}=-\frac{N}{N-1}<0$ the condition $c_{d}^{2}>0$ will
hold for
$$w_{0}<-\frac{kc^{2}+H_{0}^{2}a_{0}^{2}\left(2k_{0}+1\right)}{3\left(kc^{2}+H_{0}^{2}a_{0}^{2}\right)}$$.

Let us put $k_{0}=0.7$ and consider the value of $\Omega_{0}$
derived from (\ref{Omega01}) for various values of $w_{0}$ and
$\frac{kc^{2}}{a_{0}^{2}H_{0}^{2}}$. The results attained by such a
calculation are given in the following tables.

1. $k=0$

\begin{center}\begin{tabular}{|c|c|}
  \hline
  $w_{0}$ & $\Omega_{0}$ \\
  \hline
  -1.01 & 0.792 \\
  -1.02 & 0.784 \\
  -1.03 & 0.777 \\
  -1.04 & 0.769 \\
  -1.05 & 0.762 \\
  -1.1 & 0.727 \\
  \hline
\end{tabular}
\end{center}

2. $k=+1$

\begin{center}\begin{tabular}{|c|c|c|c|}
  \hline
  $w_{0}$ & $\Omega_{0}$ for $H_{0}^{2}a_{0}^{2}=c^{2}$ & $\Omega_{0}$ for $H_{0}^{2}a_{0}^{2}=5c^{2}$ & $\Omega_{0}$ for $H_{0}^{2}a_{0}^{2}=10c^{2}$ \\
  \hline
  -1.01 & 1.122 & 0.858 & 0.825 \\
  -1.02 & 1.111 & 0.850 & 0.817 \\
  -1.03 & 1.100 & 0.841 & 0.809 \\
  -1.04 & 1.090 & 0.833 & 0.801 \\
  -1.05 & 1.079 & 0.825 & 0.794 \\
  -1.1 & 1.030 & 0.788 & 0.758 \\
  \hline
\end{tabular}
\end{center}

3. $k=-1$

\begin{center}\begin{tabular}{|c|c|c|c|}
  \hline
  $w_{0}$ & $\Omega_{0}$ for $H_{0}^{2}a_{0}^{2}=c^{2}$ & $\Omega_{0}$ for $H_{0}^{2}a_{0}^{2}=5c^{2}$ & $\Omega_{0}$ for $H_{0}^{2}a_{0}^{2}=10c^{2}$ \\
  \hline
  -1.01 & 0.462 & 0.726 & 0.759 \\
  -1.02 & 0.458 & 0.719 & 0.752 \\
  -1.03 & 0.453 & 0.712 & 0.744 \\
  -1.04 & 0.449 & 0.705 & 0.737 \\
  -1.05 & 0.444 & 0.698 & 0.730 \\
  -1.1  & 0.424 & 0.667 & 0.697 \\
  \hline
\end{tabular}
\end{center}

\section{Integration procedure}

We have

\begin{equation}\label{fr11}
\frac{\dot{a}^{2}}{a^{2}}=\frac{8\pi
G}{3}\left(\frac{\beta^{-6/\epsilon}}{\left(a_{f}^{\epsilon}-a^{\epsilon}\right)^{3/\epsilon}}+\frac{c_{d}^{2}}{a^{3}}-\frac{3kc^{2}}{8\pi
Ga^{2}}\right).
\end{equation}
Let us consider the change of variables
$a^{\epsilon}=\xi_{f}\sin^{2}\eta$, where $\xi_{f}=a_{f}^{\xi}$,
$\eta=\eta\left(t\right)$. Then, because

$$\beta^{-6/\epsilon}=\frac{a_{0}^{3}H_{0}^{2}}{8\pi G}\frac{|w_{0}|}{|w_{0}+1|^{3/\epsilon}}\left(2k_{0}+1+\frac{kc^{2}}{H_{0}^{2}a_{0}^{2}}\right),$$

$$c_{d}^{2}=\frac{a_{0}^{3}H_{0}^{2}}{8\pi
G}\left(\frac{\left(3w_{0}+1\right)}{w_{0}}\left(1+\frac{kc^{2}}{H_{0}^{2}a_{0}^{2}}\right)+\frac{2k_{0}}{w_{0}}\right),$$

$$a_{0}=\xi^{1/\epsilon}\sin^{2/\epsilon}\eta_{0},$$

$$\sin^{2}\eta_{0}=1/N_{0}=\frac{w_{0}+1}{w_{0}},$$

Eq. (\ref{fr11}) after simplifications results  in

\begin{equation}\label{fr12}
\frac{2\sqrt{3}}{\epsilon
H_{0}}\left(\frac{w_{0}}{w_{0}+1}\right)^{3/2\epsilon}\int\frac{\left(\cos\eta\right)^{1+3/\epsilon}\left(\sin\eta\right)^{-1+3/\epsilon}d\eta}{\sqrt{\mu\left(w_{0},a_{0},\epsilon\right)\sin^{6/\epsilon}\eta+\nu\left(w_{0},a_{0},\epsilon\right)\cos^{6/\epsilon}\eta+\gamma\left(w_{0},a_{0},\epsilon\right)\sin^{2/\epsilon}\eta\cos^{6/\epsilon}\eta}}=\int
dt,
\end{equation}

where we have introduced the following three parametric coefficients

$$\mu\left(w_{0},a_{0},\epsilon\right)=\frac{|w_{0}|}{|w_{0}+1|^{3/\epsilon}}\left(2k_{0}+1+\frac{kc^{2}}{H_{0}^{2}a_{0}^{2}}\right),$$

$$\nu\left(w_{0},a_{0},\epsilon\right)=\frac{3w_{0}+1}{w_{0}}\left(1+\frac{kc^{2}}{H_{0}^{2}a_{0}^{2}}\right)+\frac{2k_{0}}{w_{0}},$$

$$\gamma\left(w_{0},a_{0},\epsilon\right)=-\frac{3kc^{2}}{H_{0}^{2}a_{0}^{2}}\left(\frac{w_{0}}{w_{0}+1}\right)^{1/\epsilon}.$$

Thus, integration of Eq. (12) from $\eta=0$ to
$\eta=\eta_{0}=\arcsin\left(\sqrt{1+1/w_{0}}\right)$ gives the age
of the universe. The lifetime of the universe can be found by
integrating Eq. (12) from $\eta=0$ to $\eta=\pi/2$. We shall in what
follows consider various possible cases.

1. Flat spacetime. The coefficients $\mu$ and $\nu$ in this case
depend only on $w_{0}$ and $\epsilon$. Coefficient $\gamma=0$. The
numerical calculations of the age of the universe and the difference
between time of final singularity $t_{f}$ and $t_{0}$ for various
values of $w_{0}$ and $\epsilon$ are given in the following table.
The used Hubble parameter for that calculation is $H_{0}=72$
km/s/Mpc.

\begin{center}\begin{tabular}{|c|c|c|c|c|c|c|}
\hline
 & \multicolumn{2}{c}{$w_{0}=-1.01$}\vline & \multicolumn{2}{c}{$w_{0}=-1.05$}\vline & \multicolumn{2}{c}{$w_{0}=-1.1$}\vline \\
\hline
  $\epsilon$ & $t_{f}-t_{0}$, Gyr & $t_{0}$, Gyr & $t_{f}-t_{0}$, Gyr & $t_{0}$, Gyr & $t_{f}-t_{0}$, Gyr & $t_{0}$, Gyr \\
\hline
  0.1 & 252.116 & 14.443 & 103.026 & 13.842 & 61.938 & 13.245 \\
  0.2 & 164.752 & 14.442 & 76.156 & 13.838 & 48.557 & 13.838 \\
  0.5 & 86.958 & 14.440 & 45.763 & 13.827 & 45.763 & 13.827 \\
  1 & 50.798 & 14.436 & 28.691 & 13.813 & 28.691 & 13.813 \\
  3 & 19.726 & 14.429 & 11.916 & 13.771 & 11.916 & 13.771 \\
  6 & 10.286 & 14.423 & 6.353 & 13.756 & 6.353 & 13.756 \\
  9 & 6.929 & 14.420 & 4.319 & 13.745 & 4.319 & 13.745 \\
\hline
\end{tabular}\end{center}

2. Spacetime with positive spatial curvature. The coefficients
$\mu$, $\nu$, $\gamma$ depend on $w_{0}$, $\epsilon$ and $a_{0}$. We
use the  values 1, 5 and 10 for the relation
$a_{0}^{2}H_{0}^{2}/c^{2}$. We obtain in this case

a)$a_{0}^{2}H_{0}^{2}/c^{2}=1$.

\begin{center}\begin{tabular}{|c|c|c|c|c|c|c|}
\hline
& \multicolumn{2}{c}{$w_{0}=-1.01$}\vline & \multicolumn{2}{c}{$w_{0}=-1.05$}\vline & \multicolumn{2}{c}{$w_{0}=-1.1$} \vline \\
  \hline
  $\epsilon$ & $t_{f}-t_{0}$, Gyr & $t_{0}$, Gyr & $t_{f}-t_{0}$, Gyr & $t_{0}$, Gyr & $t_{f}-t_{0}$, Gyr & $t_{0}$, Gyr \\
  \hline
  0.5 & 74.752 & 11.335 & 39.815 & 10.941 & 27.319 & 10.535 \\
  1 & 44.344 & 11.333 & 25.400 & 10.931 & 18.107 & 10.518 \\
  3 & 18.017 & 11.327 & 11.000 & 10.905 & 8.105 & 10.471 \\
  6 & 9.734 & 11.323 & 6.025 & 10.885 & 4.476 & 10.436 \\
  9 & 6.671 & 11.300 & 4.145 & 10.874 & 3.089 & 10.420 \\
  \hline
\end{tabular}\end{center}

b)$a_{0}^{2}H_{0}^{2}/c^{2}=5$.

\begin{center}\begin{tabular}{|c|c|c|c|c|c|c|}
\hline
& \multicolumn{2}{c}{$w_{0}=-1.01$}\vline & \multicolumn{2}{c}{$w_{0}=-1.05$}\vline & \multicolumn{2}{c}{$w_{0}=-1.1$} \vline \\
  \hline
  $\epsilon$ & $t_{f}-t_{0}$, Gyr & $t_{0}$, Gyr & $t_{f}-t_{0}$, Gyr & $t_{0}$, Gyr & $t_{f}-t_{0}$, Gyr & $t_{0}$, Gyr \\
  \hline
  0.5 & 83.422 & 12.335 & 43.857 & 11.932 & 29.912 & 11.518 \\
  1 & 48.680 & 12.333 & 27.454 & 11.923 & 19.471 & 11.501 \\
  3 & 18.825 & 12.327 & 11.341 & 11.899 & 8.359 & 11.459 \\
  6 & 9.763 & 12.324 & 6.013 & 11.883 & 4.491 & 11.429 \\
  9 & 6.550 & 12.288 & 4.073 & 11.873 & 3.060 & 11.413 \\
  \hline
\end{tabular}\end{center}

c)$a_{0}^{2}H_{0}^{2}/c^{2}=10$.

\begin{center}\begin{tabular}{|c|c|c|c|c|c|c|}
\hline
& \multicolumn{2}{c}{$w_{0}=-1.01$}\vline & \multicolumn{2}{c}{$w_{0}=-1.05$}\vline & \multicolumn{2}{c}{$w_{0}=-1.1$} \vline \\
  \hline
  $\epsilon$ & $t_{f}-t_{0}$, Gyr & $t_{0}$, Gyr & $t_{f}-t_{0}$, Gyr & $t_{0}$, Gyr & $t_{f}-t_{0}$, Gyr & $t_{0}$, Gyr \\
  \hline
  0.5 & 85.097 & 13.177 & 44.749 & 12.700 & 30.527 & 12.216 \\
  1 & 49.668 & 13.174 & 28.022 & 12.689 & 19.881 & 12.197 \\
  3 & 19.227 & 13.168 & 11.594 & 12.622 & 8.550 & 12.148 \\
  6 & 9.989 & 13.164 & 6.159 & 12.643 & 4.603 & 12.114 \\
  9 & 6.711 & 13.125 & 4.177 & 12.620 & 3.139 & 12.086 \\
  \hline
\end{tabular}\end{center}

3. Spacetime with negative spatial curvature. The coefficients
$\mu$, $\nu$, $\gamma$ depend on $w_{0}$, $\epsilon$ and $a_{0}$.
One notes that the right hand side of Eq. (\ref{fr11}) is negative
for $a_{0}^{2}H_{0}^{2}/c^{2}=1$ along the interval
$-1.1<w_{0}<-1.01$ because we consider the values 5 and 10
forrelation $a_{0}^{2}H_{0}^{2}/c^{2}$ We get

a)$a_{0}^{2}H_{0}^{2}/c^{2}=5$.

\begin{center}\begin{tabular}{|c|c|c|c|c|c|c|}
\hline
& \multicolumn{2}{c}{$w_{0}=-1.01$}\vline & \multicolumn{2}{c}{$w_{0}=-1.05$}\vline & \multicolumn{2}{c}{$w_{0}=-1.1$} \vline \\
  \hline
  $\epsilon$ & $t_{f}-t_{0}$, Gyr & $t_{0}$, Gyr & $t_{f}-t_{0}$, Gyr & $t_{0}$, Gyr & $t_{f}-t_{0}$, Gyr & $t_{0}$, Gyr \\
  \hline
  0.5 & 90.991 & 18.850 & 47.949 & 17.462 & 32.756 & 16.261 \\
  1 & 53.224 & 18.842 & 30.118 & 17.435 & 21.405 & 16.219 \\
  3 & 20.775 & 18.826 & 12.592 & 17.373 & 9.305 & 16.119 \\
  6 & 10.907 & 18.817 & 6.762 & 17.334 & 5.060 & 16.054 \\
  9 & 7.385 & 18.800 & 4.617 & 17.241 & 3.471 & 15.943 \\
  \hline
\end{tabular}\end{center}

b)$a_{0}^{2}H_{0}^{2}/c^{2}=10$.

\begin{center}\begin{tabular}{|c|c|c|c|c|c|c|}
\hline
& \multicolumn{2}{c}{$w_{0}=-1.01$}\vline & \multicolumn{2}{c}{$w_{0}=-1.05$}\vline & \multicolumn{2}{c}{$w_{0}=-1.1$} \vline \\
  \hline
  $\epsilon$ & $t_{f}-t_{0}$, Gyr & $t_{0}$, Gyr & $t_{f}-t_{0}$, Gyr & $t_{0}$, Gyr & $t_{f}-t_{0}$, Gyr & $t_{0}$, Gyr \\
  \hline
  0.5 & 88.946 & 16.241 & 46.853 & 15.381 & 31.997 & 14.564 \\
  1 & 52.009 & 16.237 & 29.412 & 15.362 & 20.893 & 14.533 \\
  3 & 20.267 & 16.226 & 12.267 & 15.317 & 9.059 & 14.457 \\
  6 & 10.613 & 16.219 & 6.568 & 15.287 & 4.913 & 14.406 \\
  9 & 7.170 & 16.203 & 4.477 & 15.212 & 3.364 & 14.316 \\
  \hline
\end{tabular}\end{center}

It can be checked that the above numerical estimates provide
realistic results for an age of the universe which weekly depends on
$\epsilon$ and $w_{0}$. The lifetime of universe, which on the
contrary strongly depends on $\epsilon$ and $w_{0}$, decreases with
growing $\epsilon$ and $w_{0}$.

\section{New solutions with big freeze singularity}

We shall consider in this section a flat universe which shows a big
freeze singularity (BFS) in its future, i.e.
$$
\rho=\beta^{-6/\epsilon}\left(a_f^{\epsilon}-a^{\epsilon}\right)^{-3/\epsilon},
$$
and
$$
p=-c^2a_f^{\epsilon}\beta^{-6/\epsilon}\left(a_f^{\epsilon}-a^{\epsilon}\right)^{-1-3/\epsilon}.
$$
Putting $\epsilon=3$, we have
$$
p=-c^2\beta^2 a_f^{3}\rho^2.
$$
In this case, one can solve the above equations to find the scale
factor. We obtain it in parametric form
\begin{equation}
\begin{array}{l}
a=a_f\left(\sin\eta\right)^{2/3},\\
t=t_f+\frac{1}{\kappa}\left(\ln|\tan\frac{\eta}{2}|+\cos\eta\right),
\end{array}
\label{solotion}
\end{equation}
where
$$
\kappa=\frac{1}{\beta}\sqrt{\frac{6\pi G}{a_f^3}},\qquad
dt=\frac{\cos^2\eta}{\kappa\sin\eta}d\eta.
$$
We set now $t_f=0$ and therefore $\eta=0$ corresponds to
$t=-\infty$, $\eta=\pi/2$ to $t=0$ (BFS) and $\eta=\pi$ to
$t=+\infty$.

Now one can see that if $\psi=a^3$ then the following equation holds
\begin{equation}
\frac{d^2\psi}{dt^2}=(v(t)-\lambda)\psi, \label{Schr}
\end{equation}
in which
$$
v=\frac{2\kappa^2\sin^2\eta(2\cos^2\eta+1)}{\cos^4\eta},\qquad
\lambda=-4\kappa^2.
$$
This is very interesting point for the spectral theory of the
Schr\"odinger equation. The potential $v(t)\to +0$ at
$t\to\pm\infty$ and $v(0)=+\infty$. But we have a bounded state
($\psi \in L^2$ and no zeros at $t\in (-\infty;+\infty)$), and this
is the case notwithstanding for which the potential has a
singularity at $t=0$ ($\eta=\pi/2$). One can check that
$$
\int_{-\infty}^{\infty} \psi^2dt=1,
$$
if $\psi=\sqrt{15\kappa/4}\sin^2\eta$.

Now we can use Eq. (\ref{Schr}) to find the second solution
$\hat\psi$ with the same potential $v$ and the same value of the
spectral parameter $\lambda$, i.e.
$$
{\hat\psi}=\psi\int\frac{dt}{\psi^2}.
$$
We get
\begin{equation}
{\hat\psi}=-2\cos\eta+\frac{4\cos\eta}{\sin^2\eta}-2\sin^2\eta\ln\left|\cot\frac{\eta}{2}\right|.
\label{sec}
\end{equation}
Therefore we have a new solution for the same expression
$$
v-\lambda=12\pi G\left(\rho-\frac{p}{c^2}\right).
$$
This solution describes two universes: the first one begins at
$t=-\infty$ and then progressively contracts until a big crunch
singularity at $t=0$ (or $\eta=\pi/2$). The second solution begins
at $t=0$ (big bang) and then starts expanding. One can see that the
Hubble root ${\hat H}=d\ln{{\hat\psi}^{1/3}}/dt$ has the asymptotic
behavior given by
$$
\lim_{t\pm\infty}{\hat H}=\pm 2\kappa,
$$
and ${\hat H}(0)=\infty$. Thus we have dS universe at
$t\to\pm\infty$.

A most interesting solution is the superposition of
$\psi=\sin^2\eta$ and ${\hat\psi}$. We can see that
$\Psi=c_1\psi+c_2{\hat\psi}$ results in a new solution
$a_{general}=\Psi^{1/3}$, such that
\begin{equation}
\frac{d^2\Psi}{dt^2}=(v(t)-\lambda)\Psi, \label{Schr1}
\end{equation}
for the same $v$ and $\lambda$. This solution describes three
distinct kinds of universes. If $c_1=1$ and $c_2=-0.1$, then we get
\newline
Universe I.  This universe begins at $t=-\infty$ and then
progressively contracts until a big crunch singularity at $t=t_i$:
$-\infty<t_i<0$ (for the case $c_1=1$, $c_2=-0.1$ we have $t_i\sim
0.72$).
\newline
Universe II.  This universe begins at $t=t_i$ (big bang at
$a_{general}(t_i)=0$) after which it starts expanding until a BFS
which takes place at $t=0$. It can be shown that this type of
universe cannot be fitted to the data of astronomical observations
and that it does not lead to a phase of accelerating expansion.
\newline
Universe III.  This universe begins at $t=0$ as a BFS (with
$a_{general}\ne 0$) and then starts expanding until it finally
behaves like De Sitter for large $t$
$$
\lim_{t\to\infty}{ H_{general}}= \frac{2\kappa}{3}.
$$

We find the latter two kinds of cosmological evolution to be most
interesting.

\section{Solutions with cosmological constant}

Let`s consider the universe filled with generalized Chaplygin gas
and nonzero vacuum energy. The equation for the scale factor
\begin{equation}\label{chapllambda}
    \frac{\dot{a}^{2}}{a^{2}}=\frac{8 \pi G}{3}(\frac{\beta^{-6/\epsilon}}{(a_{f}^{\epsilon}-a^{\epsilon})^{3/\epsilon}}+\Lambda)
\end{equation}
can be conveniently rewritten in terms of the rescaled variables
$A=a/a_{f}$ and $T=(8\pi G|\Lambda|/3)^{1/2}t$. If, without loss of
generality we set $\beta^{-6/\epsilon}a_{f}^{-3}|\Lambda|^{-1/2}=1$,
we fanally derive from (\ref{chapllambda}) an equation for A,
$$
\frac{{A^{'}}^{2}}{A^{2}}=\frac{1}{(1-A^{\epsilon})^{3/\epsilon}}+\frac{|\Lambda|}{\Lambda}
,
$$
with the prime denoting derivation with respect to $T$. The solution
for this equation can be now given in parametric form:

a)For a universe starting at a big bang
\begin{equation}
\begin{array}{l}
A=(\sin\eta)^{2/\epsilon},\\
T=T_{f}-2{\epsilon}^{-1}\int^{\pi/2}_{\eta} d\eta
\cot\eta(\cos\eta)^{3/\epsilon}(1\pm(\cos\eta)^{6/\epsilon})^{-1/2},
\end{array}
\label{solotion}
\end{equation}
where the signs "+" and "-" correspond to positive and negative
cosmological constant, respectively. We put $T_{f}=0$ and therefore
$\eta=0$ corresponds to $T=-\infty$ (big band) and $\eta=\pi/2$ to
$T=0$ (BFS).

b) For a universe starting at a BFS
\begin{equation}
\begin{array}{l}
A=(\sin\eta)^{2/\epsilon},\\
T=T_{f}+2{\epsilon}^{-1}\int^{\eta}_{\pi/2} d\eta
\cot\eta(\cos\eta)^{3/\epsilon}(1\pm(\cos\eta)^{6/\epsilon})^{-1/2},
\end{array}
\label{solotion}
\end{equation}
Setting $T_{f}=0$ and therefore $\eta=\pi/2$ corresponds to $T=0$
(BFS) and $\eta=\pi$ to $T=-\infty$ (big chrunch).

\section{Summary and conclusion}
This paper deals with several new cosmic solutions, which show or do
not show a future big freeze singularity, derived from a universe
filled with generalized Chaplygin in the cases where: (i) dark
energy is the sole component of the universal energy, (ii) the
universe contains in addition some amount of dark matter, and (iii)
the universe is also equipped with a cosmological constant. We
discuss the observational feasibility  of such theoretical
solutions, reaching the conclusion that some nonzero amount of dark
matter should be present in order for the given model with future
big freeze singularity to be compatible with current observations.

It is worth noticing that the above conclusion and the alluded
results are only valid in a classical framework. It could well be
expected that our classical treatment would break down in the
neighborhood of the singularity where the energy density tends to
infinity. Actually, a quantum-gravity consideration which
contemplated the Planck length as the ultimate resolution limit
would smooth out such a singularity and allowed for a further
evolution for the universe.

\bibliography{apssamp}% Produces the bibliography via BibTeX.

\end{document}